\documentclass[prl,floatfix,twocolumn]{revtex4}
\usepackage{times,amssymb,amsmath,ulem}
\usepackage{graphicx}
\usepackage[pdf]{pstricks}

\newcommand{\bfq}{\ensuremath{\boldsymbol{q}}}
\newcommand{\bfr}{\ensuremath{\boldsymbol{r}}}
\newcommand{\bfR}{\ensuremath{\boldsymbol{R}}}

\begin{document}

\title{Potential energy surfaces for electron dynamics modeled by floating and breathing Gaussian wave packets with valence-bond spin-coupling: An analysis of high-harmonic generation spectrum}

\author{Koji Ando\footnote{E-mail: ando\_k@lab.twcu.ac.jp}}
\affiliation{Department of Information and Sciences, 
    Tokyo Woman's Christian University, 
    2-6-1 Zenpukuji, Suginami-ku, Tokyo 167-8585, Japan}

\date{\today}

\begin{abstract}
A model of localized electron wave packets (EWPs), floating and breathing Gaussians with non-orthogonal valence-bond spin-coupling, 
is applied to compute the
high-harmonic generation (HHG) spectrum from a LiH molecule induced by an intense laser pulse.
The characteristic features of the spectrum,
a plateau up to 50 harmonic-order and a cut-off,
agreed well with those from the previous time-dependent
complete active-space self-consistent-field calculation 
[T. Sato and K. L. Ishikawa, Phys. Rev. A \textbf{91}, 023417 (2015)].
In contrast with the conventional molecular orbital picture in which the Li 2s and H 1s atomic orbitals are strongly mixed, the present calculation indicates that an incoherent sum of responses of single electrons reproduces the HHG spectrum, in which the contribution from H 1s electron dominates the plateau and cut-off, whereas the Li 2s electron contributes to the lower frequency response. The results are comprehensive in terms of the shapes of single-electron potential energy curves constructed from the localized EWP model.
\end{abstract}

\maketitle

\section{Introduction}

Electron dynamics in molecules is an emerging field of research
driven by recent advances of attosecond time-resolved laser spectroscopies
\cite{Bucksbaum07,Kling08,Krausz09,Ramasesha16}.
In particular, the high-harmonic generation (HHG) spectra induced by
intense laser field has been a major subject
\cite{Krause92,Schafer93,Corkum93,Baggesen11,Bandrauk13}
vitalized by a proposed possibility of self-probing molecules by their own
electrons, which aims as far as to probe electronic
wave functions (more precisely the Dyson orbitals) via the so-called 
molecular orbital (MO) tomography
\cite{Itatani04,Haessler11,Salieres12,Offenbacher15}.

Theoretical studies of HHG spectra
with quantum dynamical calculations of
realistic systems
have been rather limited to
small atoms and molecules such as
H, He, H$_2^{+}$, H$_2$, HeH$^{+}$, D$_3^{+}$, and LiH
\cite{Grossmann13,Geppert08,Takemoto11,Remacle11,Lotstedt12,Tolstikhin13,Li16}.
Treating more electrons in larger and more complex molecules
seems too demanding at present
unless invoking 
the time-dependent mean-field approximations 
of various levels
\cite{Schlegel07,Nest08,Yonehara08,Kato12,Haxton12,Ohmura14,Sato15}
or the density functional theory (DFT)
\cite{Fowe10,Kapoor13,Telnov13}.
The conventional 
MO and DFT 
calculations 
are based on atomic orbitals (AOs)
that are clamped at nuclear centers, with the time-dependence 
carried by the coefficients of MO or 
the configuration-interaction.
They essentially rely on the concept of one-electron orbitals
in the mean-field
that are normally delocalized over the molecule
according to its spatial symmetry.
To describe the dynamics of delocalized wave functions by
spatially fixed basis functions,
those of large wave numbers or high angular momenta are needed.

To obtain an alternative perspective, 
we have been studying a model of localized electron
wave packets (EWPs)
with non-orthogonal valence-bond (VB) spin-coupling
\cite{Ando09,Ando12,Ando16,Ando17}.
It was originally developed for a polarizable and reactive
force-field model
in condensed phase simulations to be combined with nuclear
wave packets for light atoms
\cite{Kim14,Kim14prb,Kim15,Kim16}.
For small molecules
such as H$_2$, LiH, BeH$_2$, CH$_2$, H$_2$O, and NH$_3$
in the ground electronic state,
the model gives
reasonably accurate potential energy surfaces
with the minimal number of EWPs \cite{Ando12}.
The accuracy is considered to come from the flexibility 
to describe the static correlation by the VB coupling,
the dynamic correlation by the EWP breathing, and the polarization
by the EWP floating.
The electronic excited states of LiH have been also examined by
quantizing the potential energy
curves constructed in the same way as in this work \cite{Ando16}.
The present work is 
an extension to study the
real-time quantum electron dynamics.
Along this line, we have reported recently the semiquantal WP dynamics
of Li 2s EWP in LiH
induced by an intense laser pulse \cite{Ando17}.
The computed HHG spectrum exhibited the intensity 
up to a hundred of harmonic order, but not the
characteristic plateau and cut-off. We considered this would be
due to the lack of quantum coherence in the simple semiquantal
dynamics of the localized EWP, 
and could be remedied by introducing the Monte Carlo
integration of coherent-state path-integral (CSPI) 
propagator \cite{Kuratsuji81,Ando14}. 
Nonetheless, for a small diatomic molecule such as
LiH, the full quantum dynamics in one-dimension is straightforward
on the effective one-electron potential energy curves constructed from the EWPs.
This will be a test for the adequacy of the model before proceeding to the 
CSPI calculations.

After an outline of the theory and computation in Sec. II,
potential energy curves for electron motion in a LiH molecule
are presented.
Single-electron quantum dynamics on these potential curves
induced by an intense laser
pulse are computed and the HHG spectra are examined.
Section IV concludes.

\section{Theory and Computation}

As in our previous reports 
\cite{Ando09,Ando12,Ando16,Ando17},
The electronic wave function is assumed to be an antisymmetrized product 
of spatial
and spin functions,
\begin{equation}
\Psi(1,\cdots,N)={\cal A}[\Phi(\bfr_{1},\cdots,\bfr_{N})\Theta(1,\cdots,N)],
\end{equation}
with the spatial part modeled by a product of one-electron functions,
\begin{equation}
\Phi(\bfr_{1},\cdots,\bfr_{N})=\phi_{1}(\bfr_{1})\cdots\phi_{N}(\bfr_{N}) .
\end{equation}
In contrast with the conventional VB methods that use
the AOs clumped at nuclear centers, 
we employ 
\lq floating and breathing\rq\
spherical Gaussian WPs of 
variable position $\bfq_i$ and width $\rho_i$,
\begin{equation}
\phi_i(\bfr)
=(2\pi\rho_{i}^{2})^{-\frac{3}{4}}
\exp[-|\bfr-\bfq_{i}|^{2}/4\rho_{i}^2] .
\label{eq:wpbasis}
\end{equation}
In our previous report \cite{Ando17}, 
$\bfq_i$ and $\rho_i$ were time-dependent,
but in this work, the EWPs are used just to construct the effective potential
energy curves along displacements of $\bfq_i$
on which the full quantum dynamics are evolved.
The spin part $\Theta(1,\cdots,N)$ consists of the spin eigenfunctions. 
In this work, we employ a single configuration of the perfect-pairing form,
\begin{equation}
\Theta =
\theta(1,2)
\theta(3,4)
\cdots\theta(N-1,N) ,
\end{equation}
with
$\theta(i,j)=(\alpha(i)\beta(j)-\beta(i)\alpha(j))/\sqrt{2}$.
The electronic energy,
$E=\langle\Psi|\hat{H}|\Psi\rangle/\langle\Psi|\Psi\rangle$,
is thus a function of the variables
$\{\bfq_i\}$ and $\{\rho_i\}$
at a given nuclear geometry $\{\bfR_I\}$.
We first optimized $\{\bfq_i\}$ and $\{\rho_i\}$ to minimize the
energy $E(\{\bfq_i\}, \{\rho_i\}; \{\bfR_I\})$,
to determine the optimal values $\{\bfq_i^{(0)}\}$ and $\{\rho_i^{(0)}\}$.
The effective potential function for the $j$-th electron ${\cal V}_j(\bfq)$ was then constructed 
by fixing all the variables other than $\bfq_j$ at the optimal values, 
\begin{align}
    {\cal V}_j(\bfq) = E(\bfq_1^{(0)}, \cdots, \bfq_{j-1}^{(0)}, \bfq, \bfq_{j+1}^{(0)}, \cdots, \bfq_N^{(0)}, 
    \nonumber
    \\
    \rho_1^{(0)}, \cdots, \rho_N^{(0)}; \{\bfR_I\}),
\label{eq:Veff}
\end{align}
on which
the time-dependent Schr\"{o}dinger equation
was numerically solved
with the effective one-electron Hamiltonian for the $j$-th electron,
\begin{equation}
    \hat{\cal H}_j = - \frac{\hbar^2}{2m} \nabla_{\bfq} + {\cal V}_j (\bfq) .
\end{equation}
It might appear that the electronic kinetic energies were double-counted 
since they have been computed in
$E=\langle\Psi|\hat{H}|\Psi\rangle/\langle\Psi|\Psi\rangle$.
However, 
those in ${\cal V}_j(\bfq)$ are constants with the fixed $\{\rho_i^{(0)}\}$
as the kinetic energy expectation for $\phi_i(\bfr)$ of Eq. (\ref{eq:wpbasis})
is $\hbar^2/(8m\rho_i^2)$.

Although the procedure is essentially a one-electron approximation under the
field of other EWPs,
it is distinct from those of MO and Kohn-Sham models.
The use of such EWP potentials
is related to the coherent-state path-integral theory 
in which the Gaussian WPs are identified as the coordinate
representation of the coherent-state basis 
and the action integral is determined by the energy expectation
with respect to the WPs \cite{Kuratsuji81,Ando14}.
It also has a technical advantage of removing the singularity of Coulomb potential
that can cause problems with the numerical grid methods
of quantum dynamical calculation.
(This problem would be a reason for the use of soft Coulomb potential of a form
    $1/\sqrt{r^2 + c}$ in many of the previous studies, including Ref. \cite{Sato15}
with which our results will be compared.)

The scheme was applied to a LiH molecule under an intense laser pulse.
The parameters were taken from Ref. \cite{Sato15} that employed the 
time-dependent complete-active-space self-consistent-field (TD-CASSCF) calculation.
The internuclear distance was fixed at 2.3 bohr.
The EWP centers $\bfq_j$ were displaced along the bond direction
and the electronic energies were calculated
to construct the potential curves ${\cal V}_{j}(\bfq)$ in one-dimension.
The electron dynamics were induced by a laser pulse with 
time-dependent electric field
\begin{equation}
    {\cal E} (t) = {\cal E}_{0}\sin(\omega_{0}t)\sin^{2}(\pi t/\tau) ,
\hspace*{1em}
0 \le t \le \tau ,
\label{eq:Efield}
\end{equation}
parallel to the bond direction.
The frequency
$\omega_{0}$ corresponds to the wavelength of 750 nm, the duration
$\tau$ is of three optical cycles, $\tau=3(2\pi/\omega_{0})\simeq7.51$
fs,
and the field intensity ${\cal E}_{0}$ is
$5.5 \times 10^{8}$ V/cm with the laser intensity $4.0 \times 10^{14}$ W/cm$^{2}$.
The length of the simulation box was taken to be 1200 bohr,
with the transmission-free absorbing potential
\cite{GonzalezLezana04}
of 120 bohr length at both ends.
The initial condition of the electronic wave function at $t=0$ was a Gaussian 
    function of the center $\bfq_j^{(0)}$ and width $\rho_j^{(0)}$,
    i.e., those optimized without the external field.
The wave functions were propagated with the Cayley's hybrid scheme \cite{Watanabe00}
with the spatial grid length of 0.2 bohr and the time-step of 0.01 au ($\sim$0.24 as).
The norm of the wave function stayed unity
with the deviation less than 10$^{-7}$ throughout the simulation.
The HHG spectra were computed from the Fourier transform of 
the dipole acceleration dynamics.

We note that the quantum dynamical calculations in this work 
are one-dimensional:
the EWPs are spherical in three-dimension, 
but they are used just to construct the effective potential curves 
    ${\cal V}_j (\bfq)$
along the bond axis.
The one-dimensional treatment is in accord with the TD-CASSCF calculation of 
Ref. \cite{Sato15} 
which we take as the reference for comparison.

\begin{figure}[t]
    \centering
    \includegraphics[width=0.46\textwidth]{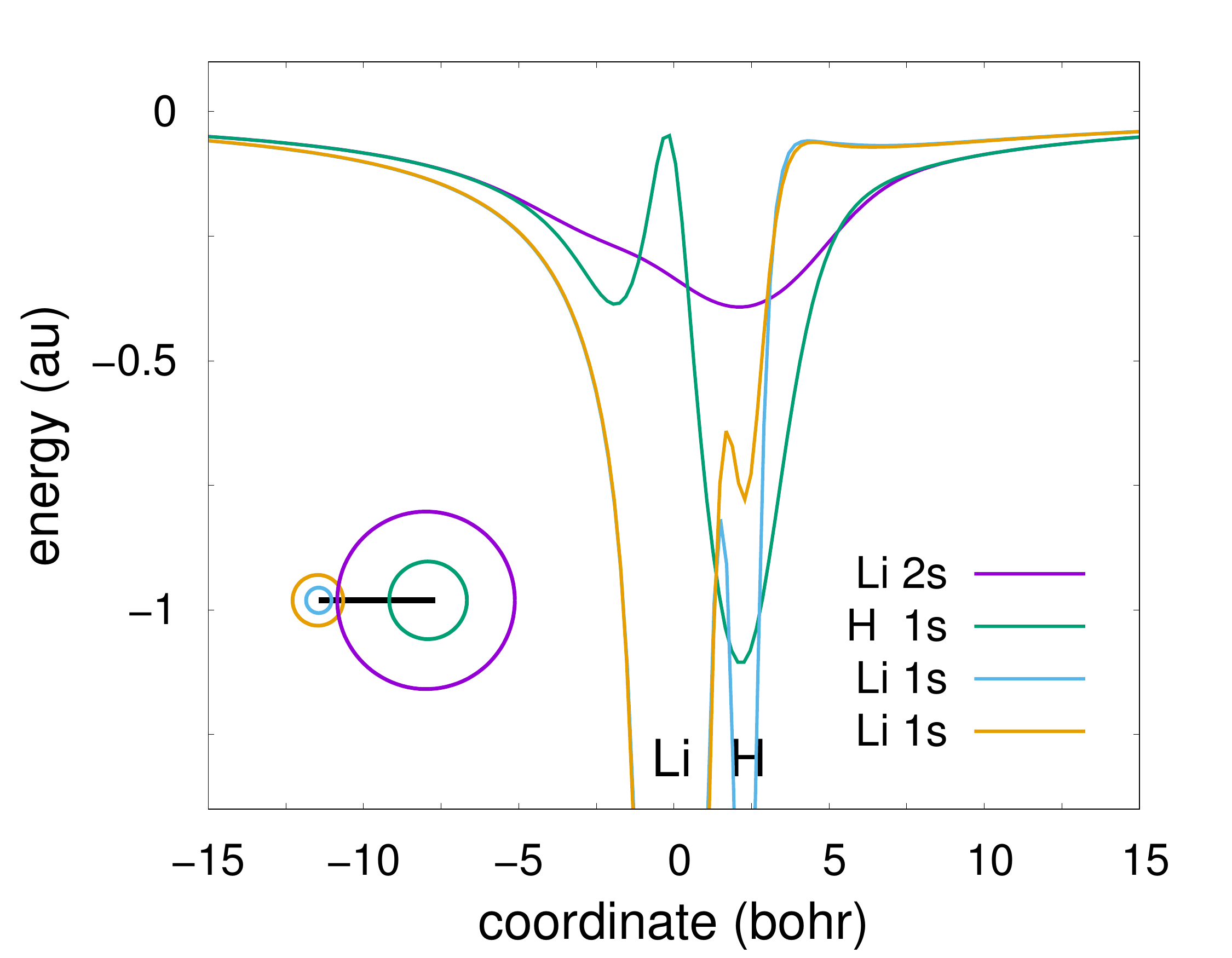}
    \caption{
        Potential energy curves for displacements of the wave packet centers
        in the singlet $X ^{1}\Sigma^{+}$ state of LiH.
        The inset shows the wave packets represented by circles
        with the radius of wave packet width $\rho_i$
    }
\end{figure}

\section{Results and Discussion}

Figure 1 displays the potential energy curves
${\cal V}_j (\bfq)$ for the electrons in LiH.
Two of them are deeply bound to the Li nuclear center
and correspond to the Li 1s core electrons. 
Their contributions to the HHG spectrum have been analyzed previously
with the ordinary
frozen-core treatment \cite{Sato15}.
Therefore, we focus on the more labile Li 2s and H 1s electrons
with much shallower potential wells in Fig. 1.
The potentials for EWPs are modulated 
under the external laser field 
via the field-dipole interaction.
The modulations at the maximum and minimum of the field 
{${\cal E}(t)$ in}
Eq. (\ref{eq:Efield})
(see also the upper panel of Fig. 3)
are displayed in Fig. 2.
The potentials indicate that the dynamics of Li 2s electron will be directly
driven by 
the laser field without energy barriers,
whereas the H 1s electron will be basically bound near the proton
but with possibilities of tunneling out in both directions.
These pictures are confirmed in Fig. 3
that plots the 
position expectation 
and its root-mean-squares (rms) deviation
of the time-dependent wave function.

\begin{figure}[t]
    \centering
    \includegraphics[width=0.46\textwidth]{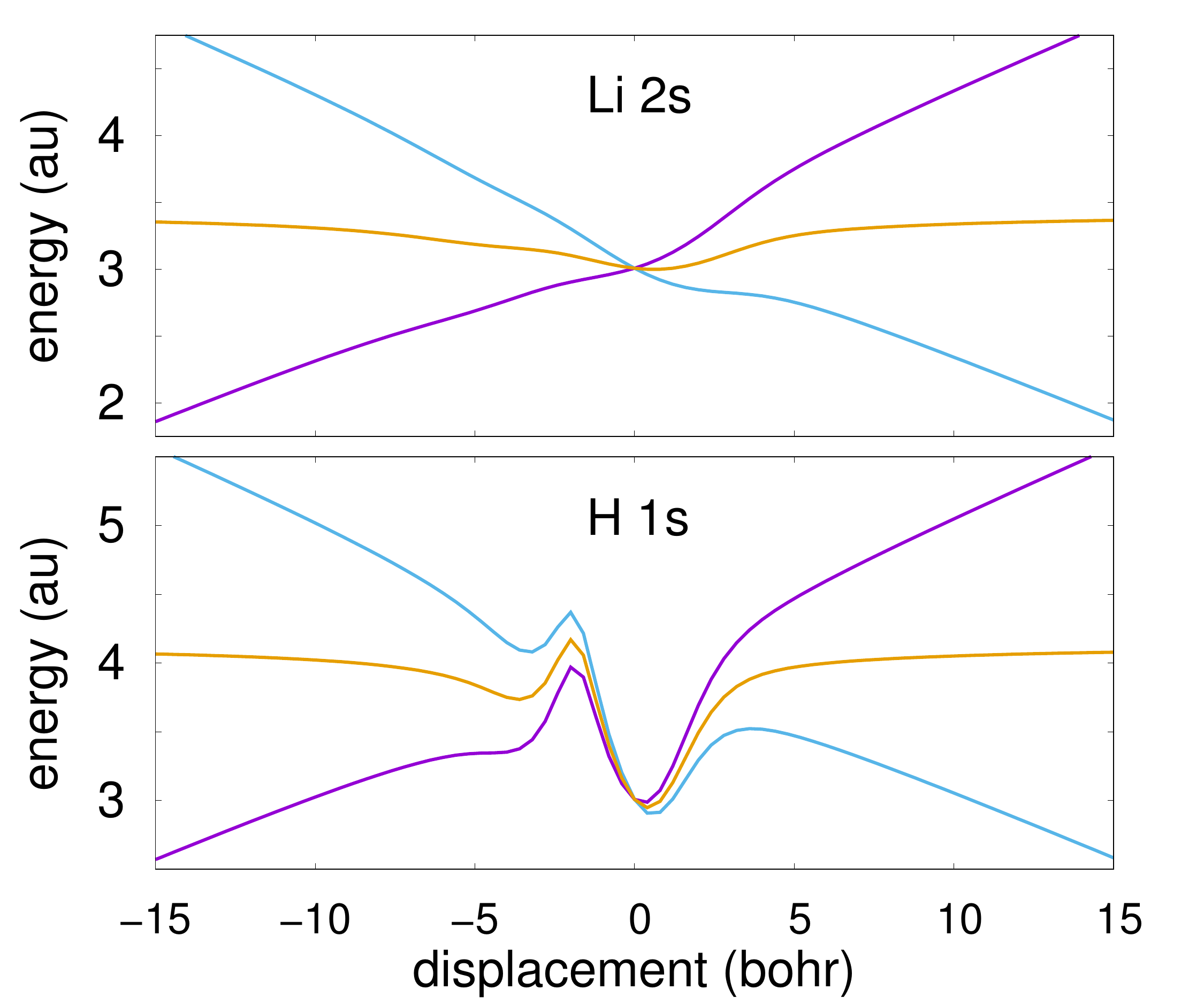}
    \caption{
        Potential energy curves for Li 2s and H 1s electron wave packet centers
        modulated by the laser field of Eq. (\ref{eq:Efield}) via the field-dipole
        interaction.
    }
\end{figure}

\begin{figure}[t]
    \centering
    \includegraphics[width=0.46\textwidth]{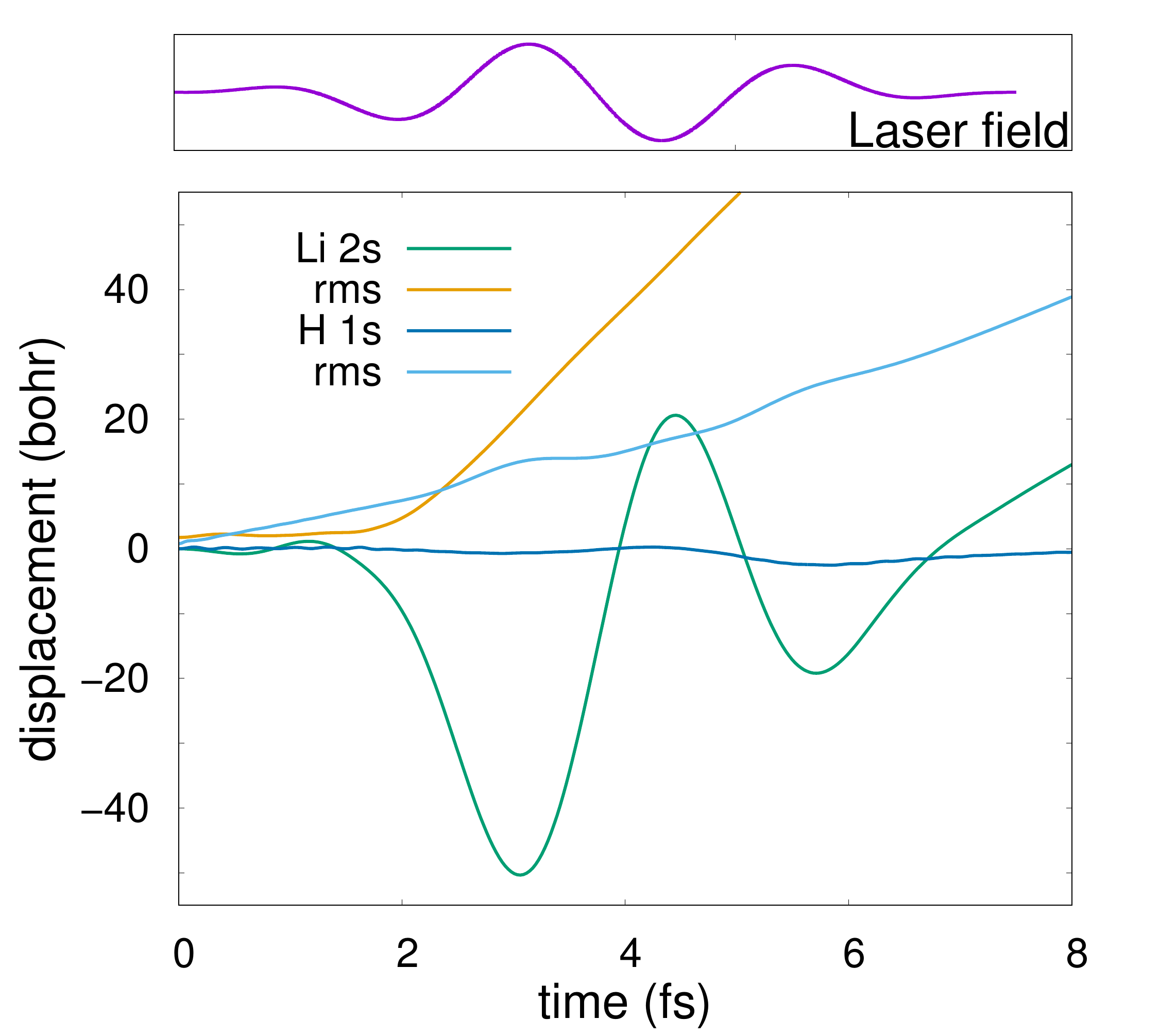}
    \caption{
        Time evolution of the position expectation value 
        and its root-mean-squares (rms) deviation.
        The upper panel displays the laser field of Eq. (\ref{eq:Efield}).
    }
\end{figure}

\begin{figure}[t]
    \centering
    \includegraphics[width=0.46\textwidth]{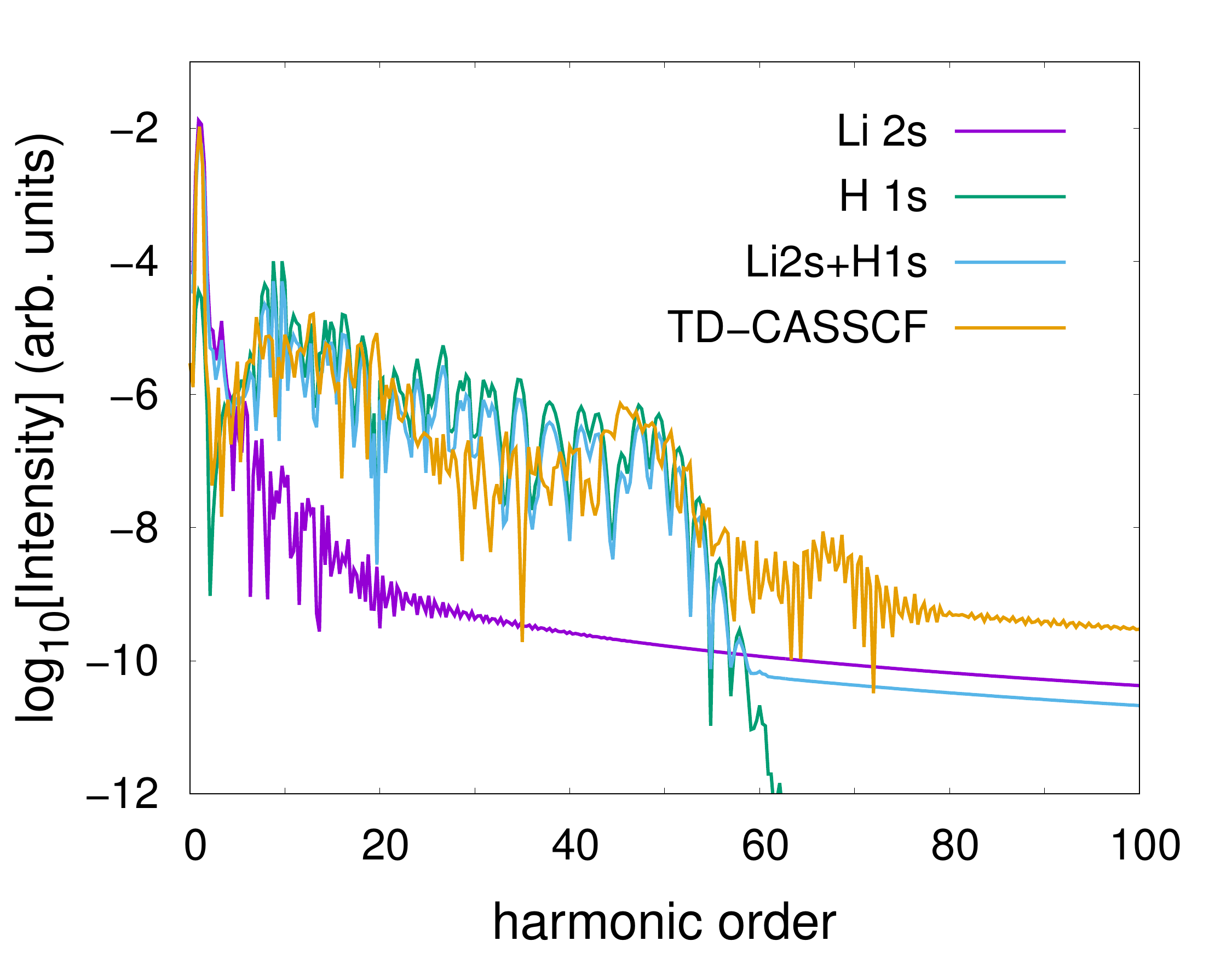}
    \caption{
        Fourier transforms of the dipole acceleration
        that give the high-harmonic generation spectra.
        The abscissa is the harmonic order $\omega/\omega_0$.
        The TD-CASSCF data is from Ref. \cite{Sato15}.
    }
\end{figure}

Figure 4 displays the HHG spectra from the 
dipole acceleration dynamics of Li 2s and H 1s electrons,
and their incoherent sum with an equal weight,
    as the dipole moment is additive,
neglecting the cross-correlation in the power spectrum.
For comparison, data from the 
TD-CASSCF calculation \cite{Sato15}
was included.
It is seen that 
the incoherent sum of the Li 2s and H 1s spectra agrees well with
the TD-CASSCF spectrum,
particularly the plateau up to $\sim$50 harmonic order (HO) and the cut-off.
The spectra from individual electrons indicate that the plateau and cut-off
come almost solely from the H 1s dynamics. 
The spectrum of Li 2s electron dominates the low-frequency peak at 1 HO,
but decays by $\sim$20 HO.
This is comprehended in terms of the potential shape and its modulation 
in the upper panel of Fig. 2:
the potential well for Li 2s EWP is shallow such that 
the dynamics will be rather similar to that of a free electron directly driven by
the external field,
which will result in the dominant contribution of the first HO in the spectrum.
The agreement of the incoherent sum with the TD-CASSCF spectrum also implies
that the correlation between the H 1s and Li 2s electrons is minor.
This is again comprehensive with the results in Fig. 3:
the small amplitude oscillation of the position expectation of H 1s electron 
indicates that the mean-field treatment for the calculation of Li 2s electronic potential
was adequate, 
and the dynamics of these two electrons with
different amplitudes of spatial oscillation are mostly decoupled. 

In the conventional MO picture, the Li 2s and H 1s AOs are strongly mixed in the
valence MOs, from which the time-dependent wave functions are described by
mixing many configuration state functions.
The picture is thus fundamentally different from the present VB WP, but
by applying some localization
or projection analysis to the complex MO-CI wave function,
a connection might be found.
Similar argument would apply to the picture of
    population transfers among MOs, in particular, the natural orbitals \cite{Kato12,Ohmura14,Brics2016}:
    in this case, a projection of the present wave functions to the
    conventional natural orbitals would complement the picture.
Another intriguing issue is on the different treatments of Coulomb interaction,
the soft Coulomb potential in Ref. \cite{Sato15} 
and the ordinary Coulomb potential in the present work.
The present scheme 
removes the singularity of $1/r$ by averaging with the EWPs $\phi_i (\bfr)$,
to give an effectively softened potential ${\cal V}_j (\bfq)$ 
without empirical adjustments.

\section{Conclusion}

The single-electron quantum dynamics on 
the electronic potential energy curves constructed with use of 
the localized floating and breathing electron wave packets with
the VB spin-coupling 
produced the HHG spectrum of a LiH molecule
in good agreement with the previous TD-CASSCF calculation.
The electronic potential energy curves provides
a unique picture for understanding the dynamics.
Although more case studies and improvements 
for numerical efficiencies are needed,
we envisage that
its applications will extend not only to other molecules
but also to
the electron conduction and optical processes in 
condensed matters.

\section*{Acknowledgment}

This work was supported by KAKENHI No. 26248009 and 26620007.
The author is grateful to Professors Takeshi Sato and Kenichi Ishikawa
for providing their data in Ref. \cite{Sato15}.

\end{document}